\documentclass[preprint]{aastex}  

\usepackage{natbib}
\usepackage{graphicx}

\DeclareGraphicsExtensions{.jpg,.pdf,.png,.eps}

\lefthead{Frieman et al.}
\righthead{SDSS Supernovae Survey.}
\shorttitle{SDSS Supernovae Survey.}
\shortauthors{Frieman et al.}

\begin{document}

\title{The Sloan Digital Sky Survey-II Supernova Survey: Technical Summary
}


\author{
Joshua~A.~Frieman,\altaffilmark{1,2,3}
Bruce~Bassett,\altaffilmark{4,5}
Andrew~Becker,\altaffilmark{6}
Changsu Choi,\altaffilmark{7}
David~Cinabro,\altaffilmark{8}
Fritz~DeJongh,\altaffilmark{1}
Darren~L.~Depoy,\altaffilmark{9}
Ben~Dilday,\altaffilmark{2,10}
Mamoru~Doi,\altaffilmark{11}
Peter~M.~Garnavich,\altaffilmark{12}
Craig~J.~Hogan,\altaffilmark{6}
Jon~Holtzman,\altaffilmark{13}
Myungshin~Im,\altaffilmark{7}
Saurabh~Jha,\altaffilmark{14}
Richard~Kessler,\altaffilmark{2,15}
Kohki~Konishi,\altaffilmark{16}
Hubert~Lampeitl,\altaffilmark{17}
John~Marriner,\altaffilmark{1}
Jennifer~L.~Marshall,\altaffilmark{9}
David~McGinnis,\altaffilmark{1}
Gajus~Miknaitis,\altaffilmark{1}
Robert~C.~Nichol,\altaffilmark{18}
Jose~Luis~Prieto,\altaffilmark{9}
Adam~G.~Riess,\altaffilmark{17,19}
Michael~W.~Richmond,\altaffilmark{20}
Roger~Romani,\altaffilmark{14}
Masao~Sako,\altaffilmark{21}
Donald~P.~Schneider,\altaffilmark{22}
Mathew~Smith,\altaffilmark{18}
Naohiro~Takanashi,\altaffilmark{11}
Kouichi~Tokita,\altaffilmark{11}
Kurt~van~der~Heyden,\altaffilmark{5}
Naoki~Yasuda,\altaffilmark{16}
Chen~Zheng,\altaffilmark{14}
Jennifer Adelman-McCarthy,\altaffilmark{1}
James Annis,\altaffilmark{1}
Roberto~J.~Assef,\altaffilmark{9}
John~Barentine,\altaffilmark{23,24}
Ralf~Bender,\altaffilmark{25,26}
Roger~D.~Blandford,\altaffilmark{14}
William~N.~Boroski,\altaffilmark{1}
Malcolm~Bremer,\altaffilmark{27}
Howard~Brewington,\altaffilmark{24}
Chris~A.~Collins,\altaffilmark{28}
Arlin~Crotts,\altaffilmark{29}
Jack~Dembicky,\altaffilmark{24}
Jason~Eastman,\altaffilmark{9}
Alastair~Edge,\altaffilmark{30}
Edmond Edmondson,\altaffilmark{18}
Edward~Elson,\altaffilmark{5}
Michael~E.~Eyler,\altaffilmark{31}
Alexei~V.~Filippenko,\altaffilmark{32}
Ryan~J.~Foley,\altaffilmark{32}
Stephan~Frank,\altaffilmark{9}
Ariel Goobar,\altaffilmark{33}
Tina~Gueth,\altaffilmark{13}
James~E.~Gunn,\altaffilmark{34}
Michael~Harvanek,\altaffilmark{24,35}
Ulrich~Hopp,\altaffilmark{25,26}
Yutaka~Ihara,\altaffilmark{11}
\v{Z}elko~Ivezi\'{c},\altaffilmark{6}
Steven~Kahn,\altaffilmark{14}
Jared~Kaplan,\altaffilmark{36}
Stephen~Kent,\altaffilmark{1,3}
William~Ketzeback,\altaffilmark{24}
Scott~J.~Kleinman,\altaffilmark{24,37}
Wolfram~Kollatschny,\altaffilmark{38}
Richard~G.~Kron,\altaffilmark{3}
Jurek~Krzesi\'{n}ski,\altaffilmark{24,39}
Dennis Lamenti,\altaffilmark{40}
Giorgos Leloudas,\altaffilmark{41}
Huan Lin,\altaffilmark{1}
Daniel~C.~Long,\altaffilmark{24}
John~Lucey,\altaffilmark{30}
Robert~H.~Lupton,\altaffilmark{34}
Elena~Malanushenko,\altaffilmark{24}
Viktor~Malanushenko,\altaffilmark{24}
Russet~J.~McMillan,\altaffilmark{24}
Javier~Mendez,\altaffilmark{42}
Christopher~W.~Morgan,\altaffilmark{9,31}
Tomoki~Morokuma,\altaffilmark{11,43}
Atsuko~Nitta,\altaffilmark{24,44}
Linda Ostman,\altaffilmark{33}
Kaike~Pan,\altaffilmark{24}
Constance~M.~Rockosi,\altaffilmark{45}
A.~Kathy~Romer,\altaffilmark{46}
Pilar~Ruiz-Lapuente,\altaffilmark{42}
Gabrelle~Saurage,\altaffilmark{24}
Katie~Schlesinger,\altaffilmark{9}
Stephanie~A.~Snedden,\altaffilmark{24}
Jesper~Sollerman,\altaffilmark{41,47}
Chris Stoughton,\altaffilmark{1}
Maximilian~Stritzinger,\altaffilmark{41}
Mark~SubbaRao,\altaffilmark{3}
Douglas~Tucker,\altaffilmark{1}
Petri~Vaisanen,\altaffilmark{5}
Linda~C.~Watson,\altaffilmark{9}
Shannon~Watters,\altaffilmark{24}
J.~Craig~Wheeler,\altaffilmark{23}
Brian Yanny,\altaffilmark{1}
and
Donald~York\altaffilmark{3,15}
}

\altaffiltext{1}{
Center for Particle Astrophysics, 
  Fermi National Accelerator Laboratory, P.O. Box 500, Batavia, IL 60510.
}
\altaffiltext{2}{
  Kavli Institute for Cosmological Physics, 
   The University of Chicago, 5640 South Ellis Avenue Chicago, IL 60637.
}
\altaffiltext{3}{
  Department of Astronomy and Astrophysics,
   The University of Chicago, 5640 South Ellis Avenue, Chicago, IL 60637.
}
\altaffiltext{4}{
Department of Mathematics and Applied Mathematics,
University of Cape Town, Rondebosch 7701, South Africa.
}
\altaffiltext{5}{
  South African Astronomical Observatory,
   P.O. Box 9, Observatory 7935, South Africa.
}
\altaffiltext{6}{
  Department of Astronomy,
   University of Washington, Box 351580, Seattle, WA 98195.
}
\altaffiltext{7}{
Department of Astronomy,
Seoul National University, Seoul, South Korea.
}
\altaffiltext{8}{
Department of Physics, 
Wayne State University, Detroit, MI 48202.
}
\altaffiltext{9}{
  Department of Astronomy,
   Ohio State University, 140 West 18th Avenue, Columbus, OH 43210-1173.
}
\altaffiltext{10}{
Department of Physics, 
University of Chicago, Chicago, IL 60637.
}
\altaffiltext{11}{
  Institute of Astronomy, Graduate School of Science,
   University of Tokyo 2-21-1, Osawa, Mitaka, Tokyo 181-0015, Japan.
}
\altaffiltext{12}{
  University of Notre Dame, 225 Nieuwland Science, Notre Dame, IN 46556-5670.
}
\altaffiltext{13}{
  Department of Astronomy,
   MSC 4500,
   New Mexico State University, P.O. Box 30001, Las Cruces, NM 88003.
}
\altaffiltext{14}{
 Kavli Institute for Particle Astrophysics \& Cosmology, 
  Stanford University, Stanford, CA 94305-4060.
}
\altaffiltext{15}{
Enrico Fermi Institute,
University of Chicago, 5640 South Ellis Avenue, Chicago, IL 60637.
}
\altaffiltext{16}{
Institute for Cosmic Ray Research,
University of Tokyo, 5-1-5, Kashiwanoha, Kashiwa, Chiba, 277-8582, Japan.
}
\altaffiltext{17}{
  Space Telescope Science Institute,
   3700 San Martin Drive, Baltimore, MD 21218.
}
\altaffiltext{18}{
  Institute of Cosmology and Gravitation,
   Mercantile House,
   Hampshire Terrace, University of Portsmouth, Portsmouth PO1 2EG, UK.
}
\altaffiltext{19}{
Department of Physics and Astronomy,
Johns Hopkins University, 3400 North Charles Street, Baltimore, MD 21218.
}
\altaffiltext{20}{
  Physics Department,
   Rochester Institute of Technology,
   85 Lomb Memorial Drive, Rochester, NY 14623-5603.
}
\altaffiltext{21}{
Department of Physics and Astronomy,
University of Pennsylvania, 203 South 33rd Street, Philadelphia, PA  19104.
}
\altaffiltext{22}{
  Department of Astronomy and Astrophysics,
   The Pennsylvania State University,
   525 Davey Laboratory, University Park, PA 16802.
}
\altaffiltext{23}{
  Department of Astronomy,
   McDonald Observatory, University of Texas, Austin, TX 78712.
}
\altaffiltext{24}{
  Apache Point Observatory, P.O. Box 59, Sunspot, NM 88349.
}
\altaffiltext{25}{
  Universitaets-Sternwarte Munich, 1 Scheinerstr, Munich, D-81679, Germany.
}
\altaffiltext{26}{
Max Planck Institute for Extraterrestrial Physics, D-85748, Garching, Munich, Germany.
}
\altaffiltext{27}{
H. H. Wills Physics Laboratory, University of Bristol, Bristol, BS8 1TL, UK.
}

\altaffiltext{28}{
Astrophysics Research Institute, Liverpool John Moores University, 
Birkenhead CH41 1LD, UK.
}
\altaffiltext{29}{
Department of Astronomy,
Columbia University, New York, NY 10027.
}
\altaffiltext{30}{
Department of Physics, University of Durham, South Road, 
Durham, DH1 3LE, UK.
}
\altaffiltext{31}{
Department of Physics, United States Naval Academy,
572C Holloway Road, Annapolis, MD 21402.
}
\altaffiltext{32}{
Department of Astronomy, University of California, Berkeley, CA 94720-3411.
}
\altaffiltext{33}{
Physics Department, Stockholm University, AlbaNova University Center, 
106 91 Stockholm, Sweden.
}
\altaffiltext{34}{
  Princeton University Observatory, Peyton Hall, Princeton, NJ 08544.
}
\altaffiltext{35}{
Lowell Observatory, 1400 Mars Hill Rd., Flagstaff, AZ 86001.
}
\altaffiltext{36}{
Jefferson Laboratory of Physics, Harvard University, Cambridge, MA 02138.
}
\altaffiltext{37}{
Subaru Telescope, 650 North A'ohoku Place, Hilo, HI 96720.
}
\altaffiltext{38}{
Institut f\"{u}r Astrophysik, Universit\"{a}t G\"{o}ttingen,
              Friedrich-Hund-Platz 1, D-37077 G\"{o}ttingen, Germany.
}
\altaffiltext{39}{
  Obserwatorium Astronomiczne na Suhorze,
   Akademia Pedagogicazna w Krakowie,
   ulica Podchor\c{a}\.{z}ych 2, PL-30-084 Krak\'{o}w, Poland.
}
\altaffiltext{40}{
Department of Physics \& Astronomy, 
San Francisco State University, San Francisco, CA 94132-4163. 
}
\altaffiltext{41}{
Dark Cosmology Centre, Niels Bohr Institute, University of Copenhagen, 
DK-2100, Denmark.
}
\altaffiltext{42}{
Department of Astronomy, University of Barcelona, Marti i Franques 1, 
E-08028 Barcelona, Spain.
}
\altaffiltext{43}{
National Astronomical Observatory of Japan, 2-21-1, Osawa, Mitaka, 
Tokyo 181-8588, Japan.
}
\altaffiltext{44}{
Gemini Observatory, 670 North A'ohoku Place, Hilo, HI 96720.
}
\altaffiltext{45}{Lick Observatory, University of California, Santa 
Cruz, CA 95064.
}
\altaffiltext{46}{
Astronomy Center, University of Sussex, Falmer, Brighton BN1 9QJ, UK.
}
\altaffiltext{47}{
Astronomy Department, Stockholm University, AlbaNova University Center, 
106 91 Stockholm, Sweden.
}



\begin{abstract}

The Sloan Digital Sky Survey-II (SDSS-II) has embarked on a multi-year project
to identify and measure light curves for 
intermediate-redshift ($0.05 < z < 0.35$) Type Ia supernovae (SNe Ia) using
repeated five-band ({\it ugriz}) imaging over an area 
of 300~sq.~deg. The survey region
is a stripe 2.5$^{\circ}$ wide centered on the celestial equator in the
Southern Galactic Cap that has been imaged numerous times in 
earlier years, enabling construction of a deep reference image for 
discovery of new objects. 
Supernova imaging observations are being acquired 
between 1 September and 30 November of 2005-7. During the first two 
seasons, each 
region was imaged on average every five nights. Spectroscopic 
follow-up observations to determine supernova type and redshift 
are carried out on a large number of telescopes. 
In its first two three-month 
seasons, the survey has discovered and measured light curves for 
327 spectroscopically confirmed SNe Ia, 30 
probable SNe Ia, 14 confirmed SNe Ib/c, 32 confirmed SNe II, plus 
a large number of photometrically identified SNe Ia, 94 of which 
have host-galaxy spectra taken so far.
This paper provides an overview of the
project and briefly describes the observations completed during the first two 
seasons of operation.

\end{abstract}

\keywords{supernovae: general, surveys}

\section{Introduction}

Type Ia supernovae (SNe~Ia) are now
well established as the method of choice for accurate relative 
distance determination over cosmological scales (see, e.g., Leibundgut 2001; 
Filippenko 2005). 
However, present cosmological constraints are based upon  
a Hubble diagram constructed from low- and high-redshift 
SN Ia samples that have been observed with a variety of 
telescopes, instruments, and photometric 
passbands. Photometric offsets between these samples are highly 
degenerate with changes in cosmological parameters and could be hidden 
in part because there is a gap or ``redshift desert'' 
between the nearby ($z\lesssim 0.1$) and distant ($z \gtrsim 0.3$) 
samples; e.g., only 6 of the 157 {\it high-quality} SN Ia light 
curves comprising the ``Gold sample'' used by Riess et al. (2004) 
fall in this gap. In addition, 
the low-redshift SN measurements used to anchor the Hubble 
diagram themselves were compiled from combinations of several nearby 
surveys using different telescopes, instruments, and selection criteria.  
While supernovae provided the first indications for the 
accelerating Universe (Riess et al. 1998; Perlmutter et al. 1999), 
obtaining precise constraints on the nature 
of the dark energy will require much improved control over such 
sources of systematic errors. Increasing precision 
calls for larger supernova samples with 
continuous redshift coverage of the Hubble diagram; it also necessitates 
{\it high-quality} data, with densely sampled, multi-band SN light curves 
and well-understood photometric calibration. 

The Sloan Digital Sky Survey-II Supernova Survey, 
one of the three components of the SDSS-II project, 
is designed to address both the paucity of SN Ia data at intermediate 
redshifts and the systematic 
limitations of previous SN~Ia samples, thereby leading to more robust 
constraints upon the properties of the dark energy (for a description 
of the SDSS, see York et al.~2000).  
With its unique combination
of large areal coverage, sensitivity, and photometric accuracy, the SDSS 
has the ability to bridge the redshift desert and aims to 
discover and measure high-quality light curves for large numbers of 
SNe~Ia at \hbox{$0.05 \lesssim z \lesssim 0.35$}.
The survey is designed to take advantage of the extensive database of
reference images, object catalogs, and photometric calibration 
previously obtained by the SDSS-I. 
The uniformity of the SDSS 
2.5-m photometric instrumentation, in conjunction with in-situ photometric 
standards {\it from the same
telescope}, minimizes systematic errors arising from 
instrumental color terms and 
multi-stage transfer of photometric standards. 
By the time it is completed, the SDSS-II SN 
survey will increase the 
number of ``fully characterized'' SN Ia
light curves (with multi-color data before peak and 
densely-sampled, well-calibrated light curves) by a significant factor.

This paper serves as an overview of the SDSS-II Supernova Survey.
The scientific motivation and survey goals are discussed in \S 2.  
The observational strategy is briefly 
described in \S 3, and \S 4 overviews the data processing and target
selection for the program. In \S 5 we describe the spectroscopic 
follow-up program, and we conclude in \S 6.
Throughout, we highlight 
some of the results from the first two seasons of observations. 
This paper serves as a companion and introduction 
to more detailed papers describing 
the methods and results from the first observing season. 
A detailed 
technical discussion of the on-mountain data processing and spectroscopic 
selection algorithm is presented by Sako et al.~(2007). 
Holtzman et al. (2007) describe the algorithm used to obtain 
SN photometry and present the photometry from the 2005 season. 
Zheng et al. (2007) describe the corresponding 
SN spectroscopy. 
Dilday et al. (2007) 
present the measurement of the SN Ia rate at low redshift from the 
2005 season. Phillips et al. (2007) and Prieto et al. (2007) present 
data on the peculiar SNe 2005hk and 2005gj. Becker et al. (2007) 
describe the discovery of trans-Neptunian objects with these data.

\section{Scientific Goals}

The SDSS-II Supernova Survey aims to address the 
following primary science goals:

\medskip
\noindent
{\bf 1. Cosmological Parameters from the 
SN Ia Hubble Diagram:} Measurements of SN Ia distances constrain the 
history of the Hubble expansion parameter. For illustration, 
assuming a spatially flat 
Universe and constant dark energy equation of state parameter $w$ (the 
ratio of its effective pressure to its energy density), the 
luminosity distance satisfies $d_L = [(1+z)/H_0] \int dz'/[
\Omega_m (1+z')^3 + (1-\Omega_m)(1+z')^{3(1+w)}]^{1/2}$, where  
$\Omega_m$ 
is the fractional density in non-relativistic matter. The SDSS-II SN survey 
will obtain distance modulus estimates for 
$\sim 100$ well-measured SNe Ia in each
of the three $\Delta z=0.1$ bins spanning the redshift range 
$0.05 < z < 0.35$.
Occupying a hitherto sparsely populated region of the Hubble diagram, 
these data 
will provide important information on the evolution of the cosmic scale
factor that is not accessible to any current supernova survey. Assuming 
that the intrinsic SN Ia distance modulus dispersion is 0.15~mag 
(in the $V$ band, Phillips et al. 1999), the SDSS-II Supernova 
Survey will determine the 
mean distance modulus with a 
{\it statistical} uncertainty of $\sim 0.015$ mag per redshift 
bin. At redshift $z=0.2$, that uncertainty is less than the difference in 
distance modulus between two cosmological models differing by 
$\delta w=0.1$ (with all other cosmological parameters fixed). 
In combination with higher-redshift SN samples from 
ESSENCE (Miknaitis et al. 2007; Wood-Vasey et al. 2007) and 
SNLS (Astier et al. 2006), and 
with other cosmological probes that primarily constrain the matter density 
$\Omega_m$ (e.g., baryon acoustic oscillations,  
Eisenstein et al. 2005), the 
SDSS-II SN Survey will lead to more robust cosmological 
constraints. The intermediate redshift range probed by the SDSS-II SN 
Survey is also of interest  
because, if $w \simeq -1$, as suggested by current data, 
dark energy began dominating over non-relativistic matter 
at $z \approx 0.4$.

\medskip
\noindent {\bf 2. Minimization and Evaluation of SN Systematics:} 
Systematic errors are comparable to statistical errors in the SN 
surveys that have reported recent results; for 
future SN cosmology studies with larger datasets, improved 
control of systematic errors will be essential. 
The systematics arising from photometric calibration 
errors are small for the SDSS:     
years of effort on the large-scale calibration of the  
imaging data (Smith et al. 2002) have achieved 1\% photometric errors over the 
area of the SDSS-II SN survey (\v{I}vezi\'{c} et al. 2007). The SDSS-II SN 
observations employ five filters, providing substantial 
color information, especially for the low-redshift portion of 
the sample. It uses a single, stable camera system with 
well-calibrated and repeatedly measured filter transmission 
curves, and the SDSS native magnitudes (which are used throughout 
this paper) are 
close to the physical 
AB system (Oke \& Gunn 1983), with well-measured offsets.
The large number of SDSS SN light curves can be partitioned into 
subsets at common redshift to investigate potential sources of 
systematic error that may afflict SN Ia distance measurements, as
well as to
search for correlations between Hubble diagram residuals
and host-galaxy morphology, metallicity, or extinction. 

\medskip
\noindent{\bf 3. Anchoring the Hubble Diagram and Light Curve Training:}
When completed, the low-redshift portion of the SDSS-II SN sample ($z\lesssim 
0.15$) 
will be large enough and of sufficiently high data quality to anchor the Hubble 
diagram and to re-train light-curve fitters, reducing or 
eliminating reliance on the  
heterogeneous low-redshift samples currently employed. 
The low-redshift portion of the SDSS sample is also 
distant enough that the effects of correlated peculiar velocities due 
to large-scale flows (Radburn-Smith, Lucey, \& Hudson 2004;
Hui \& Greene 2006; Cooray \& Caldwell 2006), 
which may be a non-negligible source of systematic 
error in existing lower-redshift samples---e.g., the ``Hubble bubble'' 
(Zehavi, et al. 1998; Jha, Riess, \& Kirshner 2007)---are small 
compared to the statistical errors. 
The large sample and high quality of supernova light curves
will provide an unprecedented opportunity to 
study in detail the color/peak brightness/decline-rate relationship, 
and to search for any additional photometric 
parameters that may further reduce the 
scatter in the Hubble diagram. 

\medskip
\noindent
{\bf 4. Rest-frame Ultraviolet Light Curve Templates For High-$z$ SN Surveys:}
SN surveys that extend to $z \gtrsim 1.5$ will be required 
to achieve robust constraints from supernovae on the time evolution of 
the dark energy equation of state (e.g., 
Frieman et al. 2003), as envisioned in the concepts proposed for the 
NASA-DOE Joint Dark Energy Mission. Hubble Space Telescope 
pilot programs in the $z>1$ regime are in progress 
(Riess et al. 2007; Barbary et al. 2006).  
To reduce systematic errors, these high-redshift SN light curves must be 
matched to a set of low-redshift templates. For example, 
at $z = 1.2$, observations in the reddest optical passbands,  
$\sim$ 8000~\AA\,, correspond to 3600~\AA\,, i.e., 
the $u$ band, in
the SN rest frame. The SDSS SNe observed  
at $z = 0.3$ will have this spectral region observed at
4700~\AA , covered by 
the SDSS $g$ band; the survey will thus improve the rest-frame
ultraviolet template data (Jha et al. 2006; Wang et al. 2005) needed for the 
interpretation of high-redshift SNe. 

\medskip
\noindent
{\bf 5. SN Cosmology from Multi-Band Photometry:}
Future wide-field imaging surveys, such as DES (Abbott et al. 2005), 
PanSTARRS (Kaiser 2002), and LSST (Tyson 2002), will 
measure optical, multi-band light curves for hundreds of 
thousands of supernovae. Spectra, which have been traditionally 
used to determine SN types and redshifts, will be obtainable for 
only a small fraction of them.  
As they evolve, SNe Ia trace out a distinct 
locus in multi-color and light-curve shape space that, 
along with host-galaxy colors,  
provides information on SN type, redshift, and age 
(Poznanski et al. 2002; Vanden Berk et al. 2001; Johnson \& Crotts 2006; 
Sullivan et al. 2006a; Sako et al. 2007). 
Future surveys 
will rely heavily on the ability to use this information to 
{\it photometrically} determine SN type and redshift. With its 
large database of multi-band SN photometry and follow-up 
spectroscopy, the SDSS SN Survey will provide an excellent 
testing ground for development of this technique.

\medskip
\noindent
{\bf 6. SN Rates, Host Galaxies, and Rare SN Types:}
In addition to the large sample of high-quality, spectroscopically confirmed
SNe~Ia, the 
survey yields light curves for a multitude of additional supernovae: 
objects that
are almost certainly SNe Ia but which we were unable to confirm 
spectroscopically, as well as other types of supernovae. Combined with 
careful monitoring of the detection efficiency using artificial SNe 
inserted into the data stream, this will enable robust measurement of 
SN rates. 
The process in which a white dwarf explodes as a SN~Ia
remains a mystery, since the progenitors are so faint that
no extragalactic event can be directly observed
before exploding. The large SDSS supernova sample
provides an unbiased census of host galaxies whose
stellar ages and metal abundances are clues to the
progenitor properties (e.g., Gallagher et al. 2005; Sullivan et al. 2006b).
Stellar population and metallicity evolution are important
systematic uncertainties in measuring dark energy with
supernovae, and their effect on SN~Ia luminosity can be
constrained with the study of a large host-galaxy sample.
The SDSS-II SN Survey will also produce a homogeneous set of Type II SN 
light curves, a small fraction of which will be spectroscopically 
identified; there is evidence 
that SNe II may also prove to be useful cosmological probes (e.g., 
Hamuy \& Pinto 2002; Baron et al. 2004). Finally, 
since it covers a larger volume than 
previous or other current SN surveys, the SDSS also probes rare 
but intrinsically interesting objects such as ``broad-line'' Type Ib/c
supernovae (some of which appear to be associated 
with GRBs; see Woosley \& Bloom 2007 for a review) 
and peculiar SNe Ia (e.g., Li et al. 2001, 2003; Hamuy et al. 2003; 
Chornock et al. 2006; Phillips et al. 2007; Aldering et al. 2006; 
Prieto et al. 2007).

\section{Observing Strategy}

The SDSS-II Supernova Survey primary instrument is the SDSS CCD camera
(Gunn et al. 1998) mounted on a
dedicated 2.5-m telescope (Gunn et al. 2006)
at Apache Point Observatory, New Mexico.  The camera obtains, nearly
simultaneously, images in five broad optical bands ($ugriz$;
Fukugita et al.~1996). The camera 
is used in time-delay-and-integrate (TDI, or drift scan) mode, which 
provides efficient sky coverage.
The Supernova Survey scans at the normal (sidereal) SDSS 
survey rate, which yields 55-s integrated exposures in each 
passband; the instrument
covers the sky at a rate of approximately 20~sq.~deg~hour$^{-1}$ and
achieves 50\% detection completeness for stellar sources at 
$u=22.5, g=23.2, r=22.6, i=21.9$, and $z=20.8$ (Abazajian et al. 2003). 
For comparison, the typical peak 
magnitude for a SN Ia with no extinction is $r \simeq 19.3, 20.8$, and 
$21.6$ mag for $z=0.1, 0.2$, and $0.3$.

The SDSS-II Supernova Survey 
scans a region (designated stripe~82) centered on the celestial equator
in the Southern Galactic hemisphere that is 
2.5$^{\circ}$ wide and runs between right ascensions of
20$^{\rm hr}$ and 4$^{\rm hr}$, 
covering a total area of~300~sq.~deg.  
Since there are gaps between the six CCD columns in the focal plane, 
this area is typically covered by
alternating, on successive nights of observation,
between the interleaving northern and southern declination strips of the
stripe.
In addition to being nearly optimal for finding SNe Ia in the 
redshift desert, sidereal scanning 
benefits from many years of SDSS survey 
operations and data processing experience. Except near its ends, the stripe 82   
area has low Galactic extinction (Schlegel et al. 1998), 
can be observed from Apache Point at low airmass 
from September through November, and is accessible from almost all
ground-based telescopes for subsequent spectroscopic and photometric 
observations. This stripe has two important additional features, arising 
from the fact that 
it has been imaged numerous times during the 
SDSS-I survey (2000-2005) in photometric conditions with good seeing 
($<1.5''$). First, 
the resulting deeper, co-added images, with a median of 10 single-epoch 
exposures per band, provide a high-quality,
photometrically calibrated template for carrying out image subtraction 
to discover supernovae. Second, the repeated 
imaging  
has enabled improved photometric calibration at the $\sim 1$\% level 
(\v{I}vezi\'{c} et al. 2007). 

\begin{figure}[!ht]
\centering
\includegraphics[width=4.in]{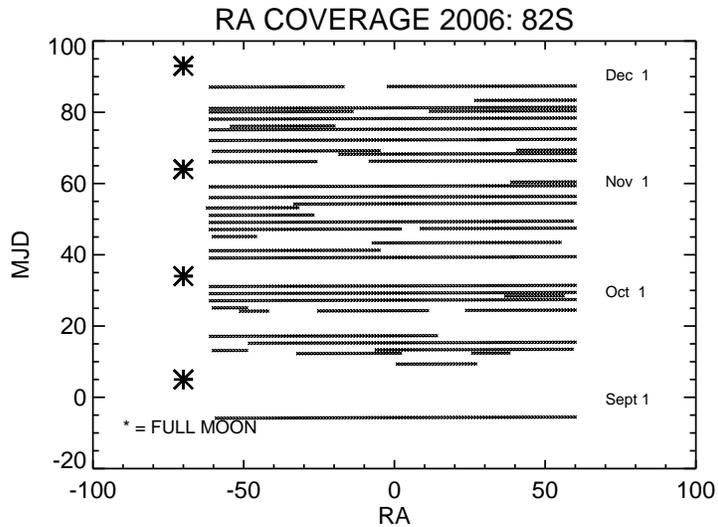}
\caption{Right Ascension coverage (in degrees) vs. time (measured from 
Sept. 1, i.e., MJD-53980) for the southern half of stripe 82 
during the 2006 SDSS SN season. 
The large asterisks denote gaps 
around full moon. The first scan was taken in late August to minimize 
survey edge effects. 
The first part of September 2006
suffered from poor observing conditions.
}
\label{fig:area}
\end{figure}

\begin{figure}[!ht]
\centering
\includegraphics[width=3.5in]{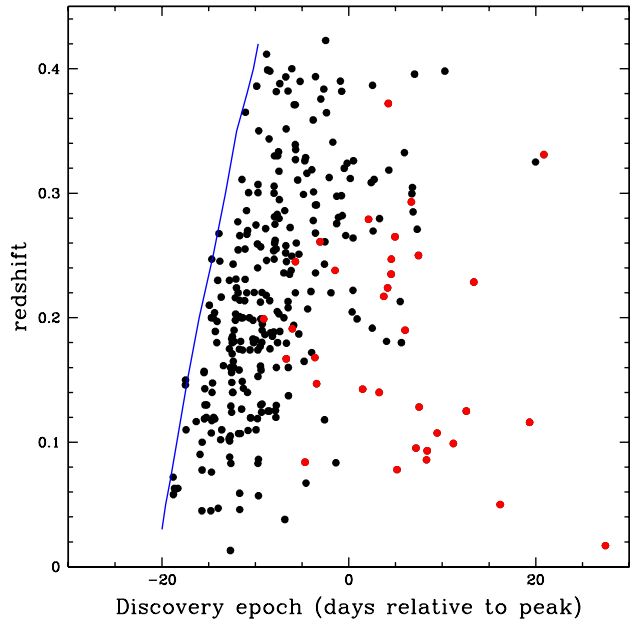}
\caption{Photometric discovery epoch relative to estimated time of 
$g$-band peak light vs. Redshift,  
for 312 spectroscopically confirmed SNe Ia from the 2005 and 2006 seasons. 
The epoch of peak light 
is determined from light curve fits to the on-mountain  
photometry. Black points denote supernovae that reached 
peak light at least seven days after the start of the survey on Sept. 1;   
red points denote supernovae that peaked before Sept. 7 and 
account for most 
of the SNe found after peak. Blue curve shows expected epoch vs. 
redshift for a fiducial model SN Ia with $r=22.5$, $\Delta m_{15}=1.2$, and no 
extinction. 
}
\label{fig:epoch}
\end{figure}

During the period 1 September through 30 November 2005-7, the 
Supernova Survey is carried out on most of the useable observing nights.   
The exceptions are the five brightest nights around each full moon, 
nights used for telescope engineering (which generally bracket 
the five full moon nights), and occasional photometric 
nights used by the SDSS-II SEGUE project for imaging. 
Accounting for weather, the Moon, and the interrupts above, each 
region in  the survey area 
has been imaged on average every five nights during the first 
two seasons, while 
the average on-sky cadence during non-bright time is about 
every 4.5 nights, as shown in 
Figure \ref{fig:area}. 
As a rolling survey with high cadence, SDSS SN is able to find  
supernovae well before maximum light (for $z\lesssim 0.3$); 
Figure \ref{fig:epoch} shows that the vast majority of spectroscopically 
confirmed SNe Ia were discovered before peak light. The discovery 
epochs shown here are from the on-mountain search photometry; in some 
cases, final photometric reductions reveal detections at even 
earlier epochs. We see that the discovery epoch vs. redshift 
limit is consistent with a 
photometric limit for SN Ia 
discovery of $r \simeq 22.5$. This survey cadence and 
detection limit result in well-sampled, 
multi-band light curves, as shown, e.g., in Figure \ref{fig:lightcurve} (see 
also Holtzman et al. 2007). 
Since SNe Ia typically spend about 25 
rest-frame days brighter than 1 mag below peak light, 
the three-month annual survey duration is long enough  
that the efficiency hit due to survey ``edge effects'' (in time) 
is acceptable. The loss can be further minimized by carrying out 
targeted photometric follow-up observations of SNe on other telescopes into 
December (see \S \ref{sec:follow}).

\section{Data Processing and Target Selection}
\label{sec:select}
We use a dedicated computer cluster at the observatory 
to perform rapid reductions of the SDSS-II SN imaging data (see Sako et al. 2007 
for more details). The $ugriz$ data are processed through the first stages of 
the standard SDSS photometric reduction pipeline (Lupton et al. 2001) to 
produce ``corrected,'' astrometrically calibrated (Pier et al. 2003) 
images for the SN search data. The deeper 
co-added reference images, comprising data taken up through 
2004, are convolved to match the point-spread functions 
of the search frames and subtracted from 
them, using a modified version of the frame subtraction pipeline 
developed for ESSENCE (Miknaitis et al. 2007). 
``Real-time'' image subtraction is limited to 
the $gri$ bands, which are the bands most useful for SN detection in 
the redshift range of interest.
The subtracted images are processed through an automated object 
detection algorithm, which yields initial photometric measurements.
Objects detected in more than one passband that are not coincident 
with previously catalogued 
stars or variable objects and not detected as moving 
during the $\sim 5$-minute 
interval between the $g$ and $r$ exposures are selected for further study. 
For a full night of 
imaging, this processing is generally completed within 24 hours or less, 
and the resulting object information is sent 
electronically to a computer at Fermilab.

\begin{figure}[!ht]
\includegraphics[width=7in]{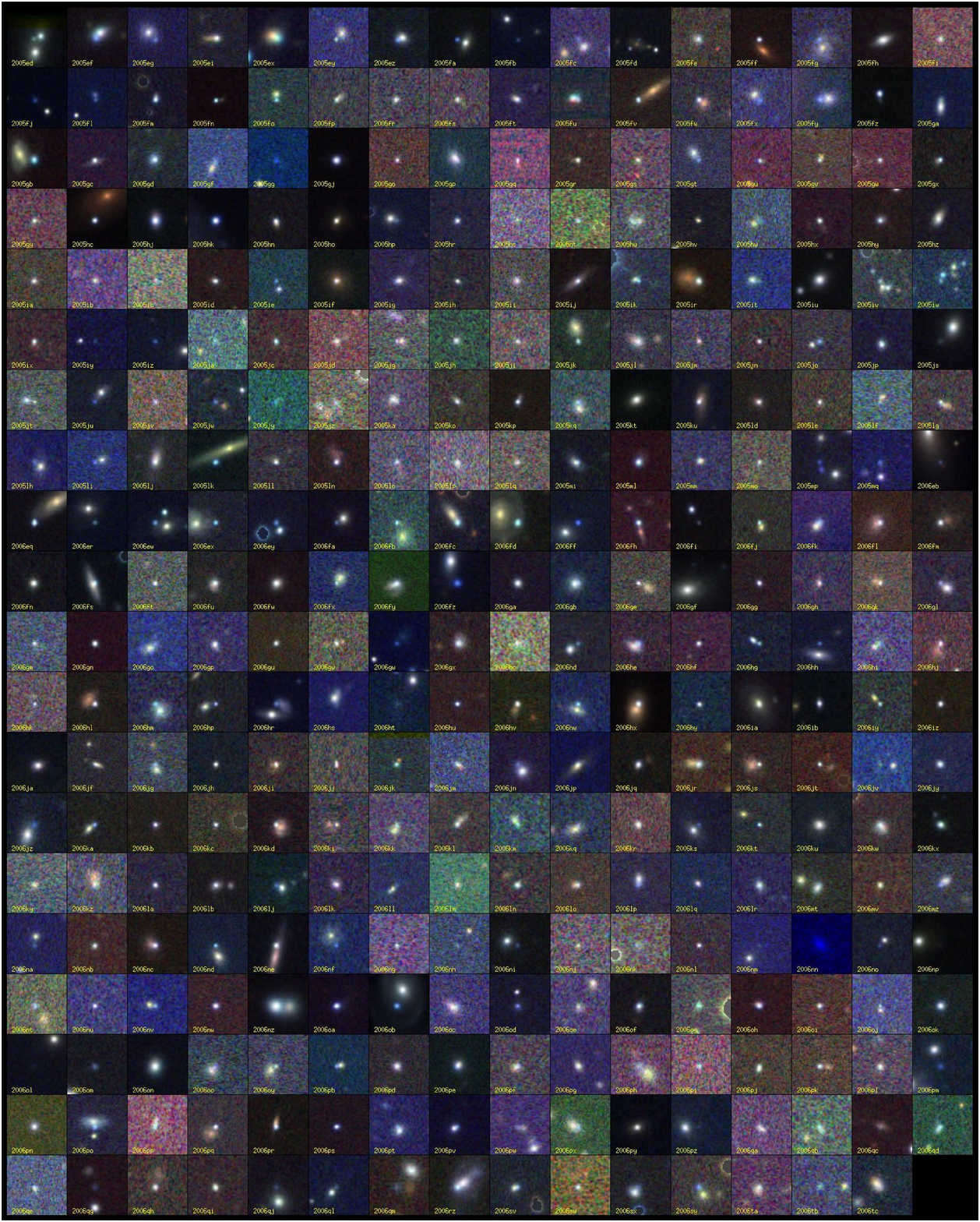}
\caption{Spectroscopically confirmed SDSS SNe Ia from 2005 and 2006.}
\label{fig:gallery}
\end{figure}

For the Fall 2005 observing season, image cutouts of all 
objects passing the above criteria were  
manually evaluated using a web interface. 
For 2006, additional and improved software cuts were applied to 
reduce the number of objects visually scanned 
without sacrificing discovery potential. In 2006, single-epoch 
detections were only passed to manual scanning if they were 
not detected as moving (with an improved ``autoscanner'' algorithm for 
vetoing moving objects; see Sako et al. 2007) and if they were bright, 
with either $r<21$ or $g<21$. Otherwise, at least two detection epochs
were required for an object to be scanned. Visual screening 
is mainly intended to reject artifacts and other objects that 
are obviously not supernovae; artificial supernovae inserted into 
the images are used to monitor the efficiency of this process. 
Objects that pass this screening are denoted 
 ``candidates''. Subsequent object detections at the 
same position are automatically added to the database of 
information about each candidate. 
In 2005, approximately 6,753 sq. deg. of imaging data were 
processed on the mountain; 
145,395 objects were manually scanned, yielding 11,385 unique 
SN candidates and 130 spectroscopically confirmed SNe Ia. In 2006, 
approximately 7,354 sq. deg. of data were processed; 
only 14,404 objects were scanned, yielding 3,694 candidates and 
197 spectroscopically confirmed SNe Ia 
\footnote{ 
The viability of detecting and following SNe from SDSS imaging data 
was demonstrated in earlier pilot runs.  
In Fall 2002, 77 SN candidates were detected in  
several SDSS scans on stripe 82; 39 of them were subjected to subsequent
spectroscopic observation, resulting in 18 confirmed SNe Ia 
out to $z \simeq 0.4$ and 8 SNe of other types (Miknaitis et al. 2002).
In Fall 2004, an engineering test run for the SDSS-II SN Survey 
was carried out 
using 20 nights scheduled over 1.5 months, yielding 16 spectroscopically 
confirmed SNe Ia, 5 SNe II, and 1 SN Ib/c (Sako et al. 2005).}.
A gallery of images of the confirmed SNe Ia is shown in Figure 
\ref{fig:gallery}.

The  $gri$ light curve for each SN candidate is 
fit with templates for different SN types and redshifts, allowing 
for variable reddening  
and, for SNe Ia, variable light-curve stretch. 
If host-galaxy spectroscopic or photometric redshift information is 
available from the SDSS database, 
it can also be included in the fit. As data on a candidate 
accumulate over time, the fits are updated. Based on relative $\chi^2$ 
values, SN Ia-like light curves are identified, and a large subset of 
them are targeted for spectroscopy to determine redshift 
and confirm SN type. All 
SN Ia candidates found before peak and with 
estimated current $r$-band magnitude $\lesssim 20$ are placed on the 
target list; they are generally accessible with 3-4 m class telescopes, 
so our follow-up observations are 
nearly complete out to that magnitude. There is a 
much larger number of fainter, higher-redshift candidates for which 
the follow-up resources are limited; those are prioritized based on 
good light-curve coverage, low estimated host-galaxy light contamination, and 
low host-galaxy dust extinction estimated from the light-curve fits. The 
basic goal is to target those SNe for which we expect to 
achieve accurate distance estimates based on the SDSS photometry.  
For details of the target selection algorithm, 
see Sako et al. (2007). 

\begin{figure}[!ht]
\centering
\plottwo{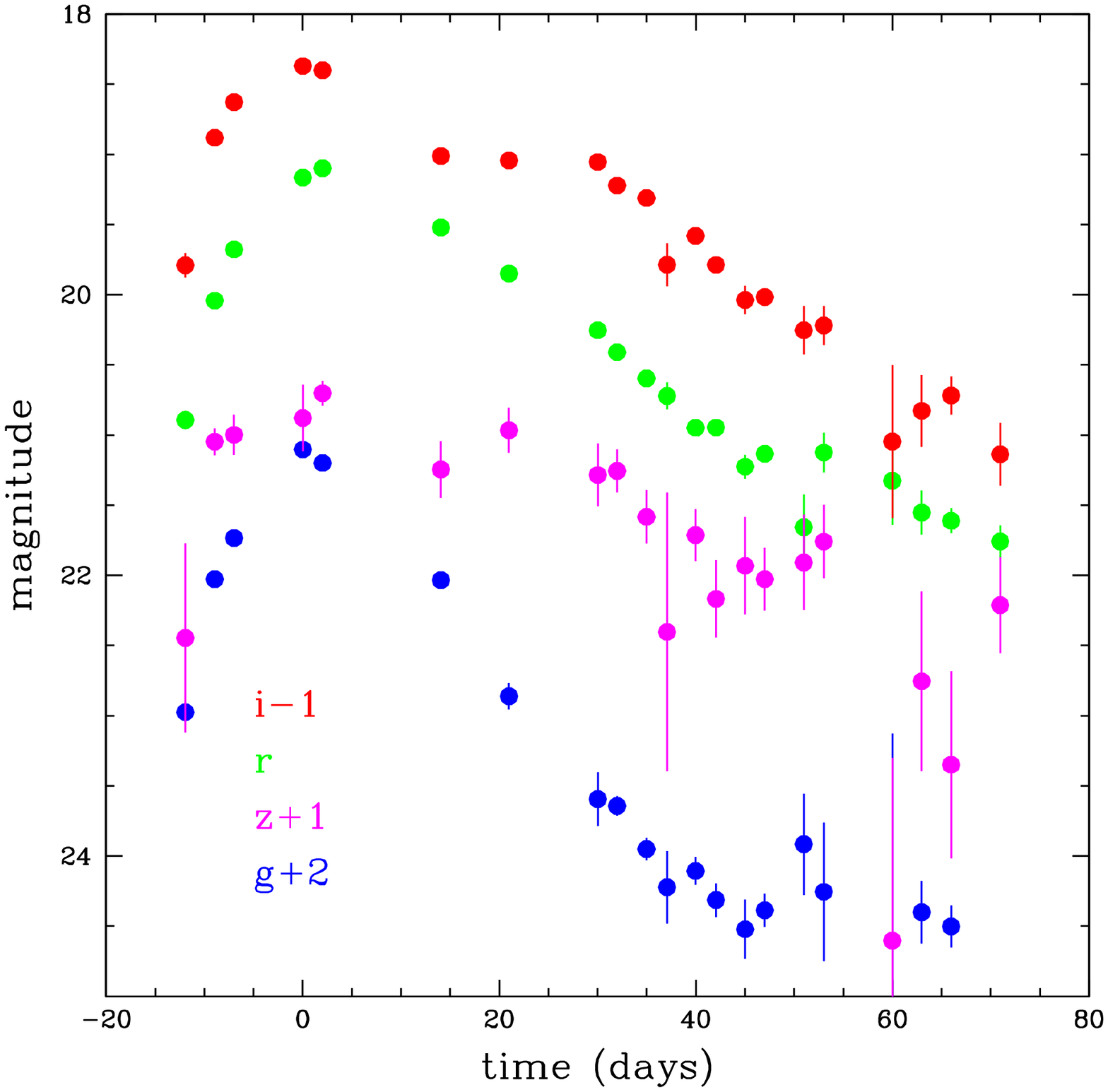}{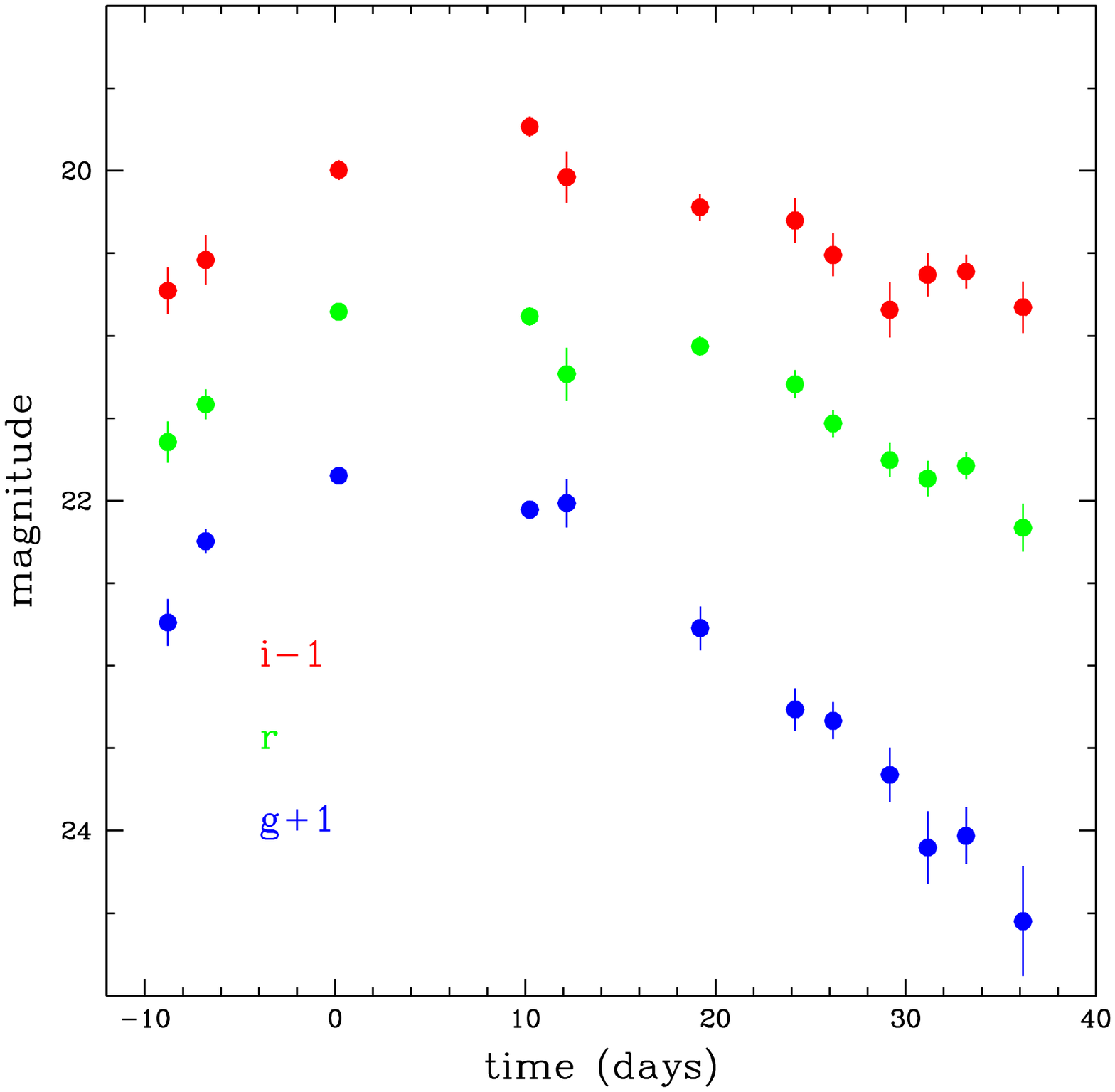}
\caption{{\it Left}: $griz$ light curves for SN 2005ff, a confirmed 
SN Ia at $z=0.088$; this 
light curve is particularly well sampled because it lies in the 
region where the northern and southern halves of stripe 82 overlap. 
{\it Right}: $gri$ light curves for SN 2005gg, a confirmed SN Ia at $z=0.231$. 
Time is measured in days from peak of $g$-band light.
}
\label{fig:lightcurve}
\end{figure}

Final SN photometry is carried out after each season 
using the ``scene modeling'' code 
developed for the SDSS-II SN Survey 
(Holtzman et al. 2007). A sequence of stars around 
each SN is taken from the list of \v{I}vezi\'{c} et al. (2007), who derived 
standard magnitudes from multiple observations taken during the 
main SDSS survey under photometric, good seeing conditions. Using these 
stars, frame scalings and astrometric solutions are derived for each 
of the SN frames, as well as for pre-supernova frames on stripe 82 
taken as part of either the main SDSS survey (pre-2005) or the SN survey. 
Finally, the entire stack of frames is simultaneously fit for a single 
supernova position, a fixed galaxy background in each filter (characterized 
by a grid of galaxy intensities), and the supernova brightness in each 
frame. Final photometry is carried 
out in all five SDSS passbands; however, since SNe Ia are intrinsically 
UV-faint and the SDSS system throughput is low in the $z$ band, 
we typically detect 
SNe Ia in the $u$ and $z$ bands only at modest redshift, $z\lesssim 0.15$.
Figure \ref{fig:lightcurve} shows light curves for two SNe Ia based 
on the outputs of the scene modeling photometry code.

\begin{figure}[!ht]
\centering
\includegraphics[width=3.5in]{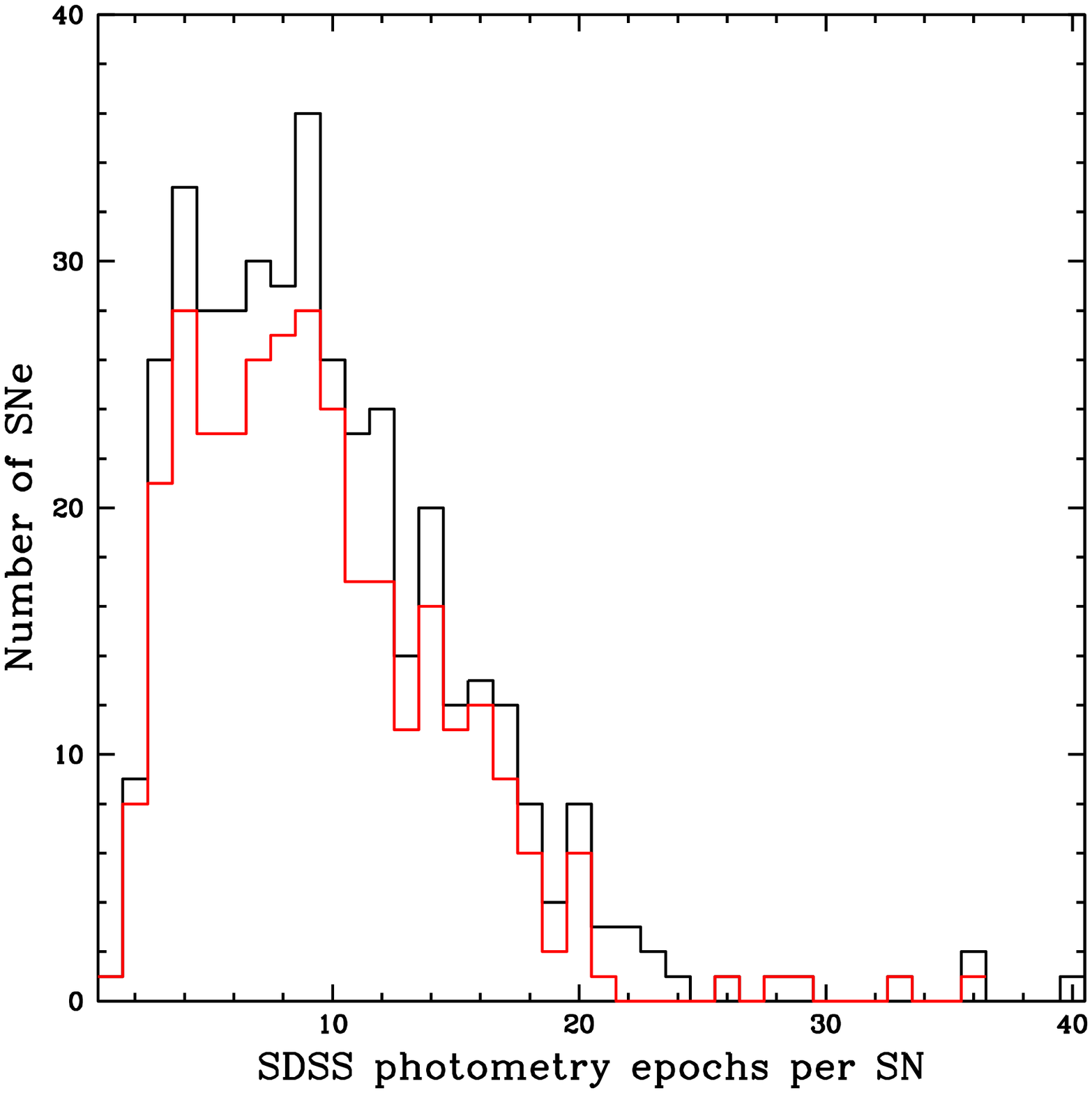}
\caption{Distribution of number of SDSS photometry epochs for 
confirmed SNe of all types (black) and for SNe Ia (red) for 
the 2005 and 2006 seasons, based on 
the on-mountain photometric reductions.
}
\label{fig:nepoch}
\end{figure}

Figure \ref{fig:nepoch} shows the approximate distribution of the number of 
SDSS photometry epochs for the spectroscopically confirmed 
SNe from the first two seasons for all types (in black) and for 
SNe Ia (in red). An epoch here constitutes 
a detection in at least two of the $gri$ bands; in the majority of 
cases, the SN is detected in all three bands. The median number of 
photometric epochs per supernova is 9 based on the on-mountain 
photometric detections. The final scene-modeling photometry 
often produces detections at additional epochs at faint flux levels, and 
additional photometric epochs observed 
with other telescopes are available for a subset of the supernovae
(see discussion at end of \S \ref{sec:follow}). 

All spectroscopically confirmed 
supernovae are announced in a timely manner through the Central 
Bureau for Electronic Telegrams. In addition, bright (i.e., 
low-redshift) 
SN candidates observed for at least two epochs and found on the rise are 
announced, often before spectroscopic confirmation. Moreover, 
as of the 2006 season, all candidates, with all relevant photometric 
information (based on the on-mountain reductions), are made available 
on the worldwide web as soon as they are 
identified\footnote{http://sdssdp47.fnal.gov/sdsssn/sdsssn.html.}. 
In collaboration with VOEventNet (Drake et al. 2006), 
as of 2006 we are also 
sending all candidates verified by a second epoch of observation to a
publicly accessible messaging system, regardless of the results of
our light-curve fits to supernova templates.  By using this rapid
messaging system and standard communication packets structured as
XML-format VOEvents, in principle 
we enable nearly instantaneous follow-up observations of our
candidates by a global network of telescopes subscribing to the
VOEventNet feed.  In 2006, we released information on 3075 two-epoch
candidate events\footnote{http://voeventnet.caltech.edu/SDSS.shtml.}, 
and we expect to generate a similar number of events during the 
2007 season. Finally, 
the corrected images and photometric catalogs for all the SN data 
are made available 
to the community soon after they are reprocessed through the SDSS 
photometric pipeline at Fermilab\footnote{  
http://www.sdss.org/drsn1/DRSN1$_{-}$data$_{-}$release.html.}.

\section{Follow-up Observations: Spectroscopy \& Photometry}
\label{sec:follow}

Spectroscopy of SN candidates is carried out 
on a number of telescopes, primarily to 
determine SN type and redshift but also to measure 
other properties. Where 
possible, a host-galaxy spectrum is extracted as well, enabling more accurate 
redshift determination from the narrower galaxy spectral features and 
determination 
of host-galaxy spectroscopic type. For our sample, 
the redshifts based on host-galaxy 
features are accurate to about $\Delta z= 0.0005$, while 
those based on SN Ia features 
are accurate to about $\Delta z=0.005$. For approximately two-thirds 
of the confirmed SNe Ia, we have redshifts determined from host-galaxy 
features.

\begin{table}[!h]
\centering
\begin{tabular}{|c|c|c|c|}
\hline 
Telescope  & Aperture (m) & Nights Scheduled & SNe Observed \\
\hline
HET & 9.2 &  64.5 hrs (2005)& 66  \\
 & & 80 hrs (2006)& 77 \\
\hline
NTT & 3.6 & 17 (2006)& 70 \\
\hline
ARC & 3.5 & 31 half-nights (2005)& 47  \\
& & 28 half-nights (2006)& 31 \\
\hline
Subaru  & 8.2 & 6 shared (2005) & 33 \\
& & 4 (2006)& 33 \\
\hline
MDM & 2.4 & 49 shared (2005)& 27 \\
& & 50 shared (2006)& 26 \\
\hline
WHT & 4.2 & 6 (2005)& 26  \\
\hline
KPNO & 4 & 6 (2006)& 15 \\
\hline
Keck & 10 & 1 ToO (2005)& 14 \\
& & 1 ToO (2006)& 4 \\
\hline
NOT & 2.6 & 4 (2006)& 10 \\
\hline
SALT & 11 & ToO (2006)& 5 \\
\hline
\end{tabular}
\caption{Telescope resources allocated 
for spectroscopic observations in 2005 and 2006. Actual 
time used is less due to weather and instrument problems.
Shown are the amounts allocated for Sept.-Dec. 
Target of 
opportunity (ToO) denotes time that was not specifically 
scheduled for SDSS SN 
observations. The HET and SALT are queue-scheduled. A portion of 
the MDM and ARC time allocation listed here was used for 
additional photometric observations. The 
last column includes multiple observations of 
the same SN and includes all confirmed SN types. Legend (for 
more information, see acknowledgements at the end of the paper): 
HET: Hobby-Eberly Telescope, McDonald Observatory, Texas; NTT: 
ESO New Technology 
Telescope, La Silla Observatory, Chile; 
ARC: Apache Point Observatory 3.5 m telescope, New 
Mexico; Subaru: Subaru Telescope, National 
Astronomical Observatory of Japan, Hawaii; MDM: Hiltner Telescope, 
MDM Observatory, 
Arizona; WHT: William Herschel Telescope, 
La Palma; KPNO: Kitt Peak National Observatory, Arizona;  
Keck: W.M. Keck Observatory, Hawaii; NOT: Nordic Optical Telescope, 
La Palma; SALT: South African Large Telescope, South African Astronomical 
Observatory. 
  }
\label{tab1}
\end{table}

Table \ref{tab1} lists the spectroscopic resources, the 
amount of time scheduled on them for each of the first two seasons, 
and the number of SNe observed. 
In addition to those listed in Table \ref{tab1}, in a handful of cases 
spectra have been taken by 
other observers/telescopes based on information we have released 
in electronic circulars or through 
informal communication. These include the Center for Astrophysics 
group (S. Blondin, M. Modjaz, R. Kirshner, P. Challis, M. Hicken, 
M. Calkins, L. Macri), 
the Nearby Supernova Factory, and the ESSENCE team.

\begin{figure}[!ht]
\centering
\epsscale{1.11}
\plottwo{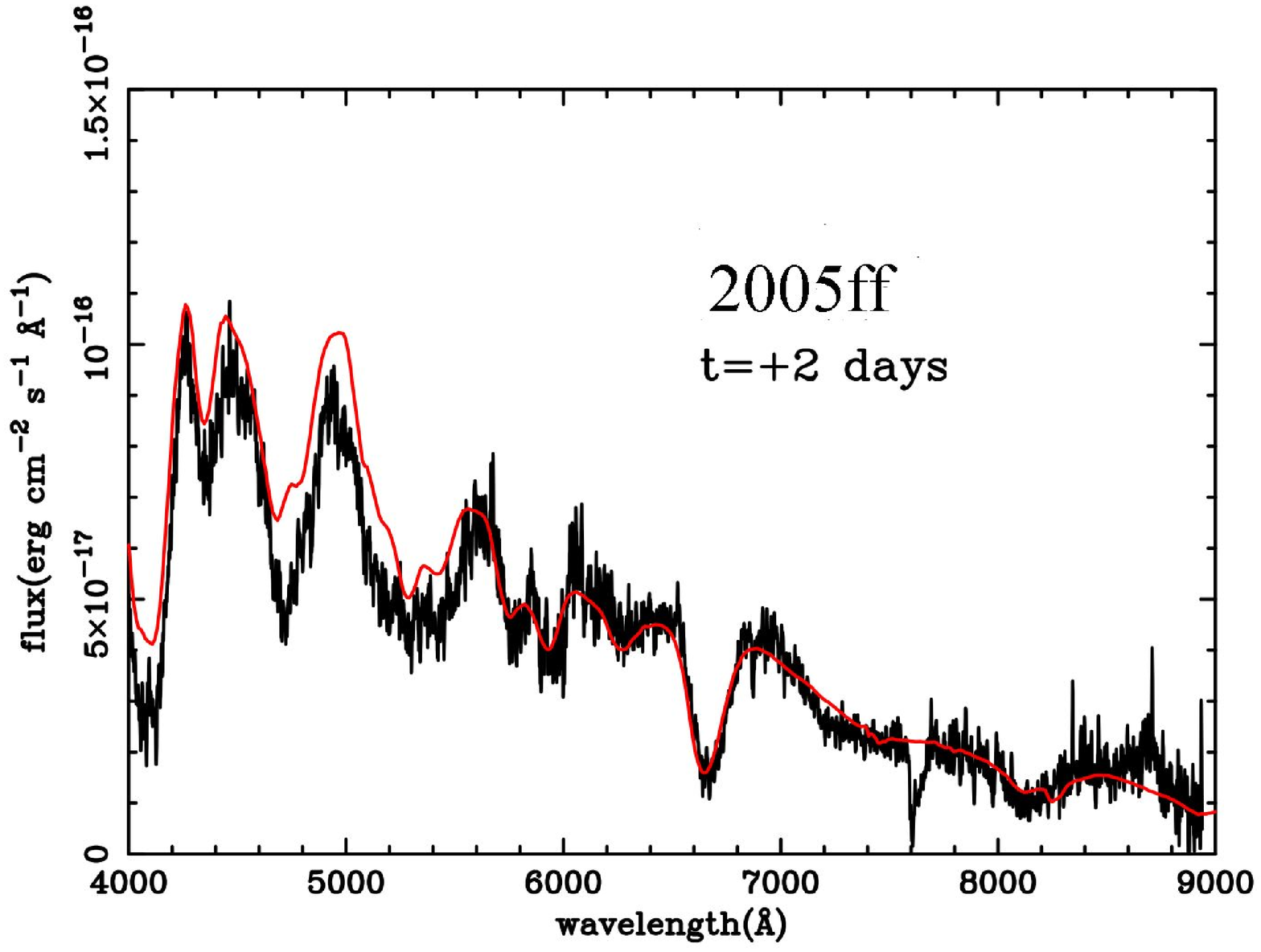}{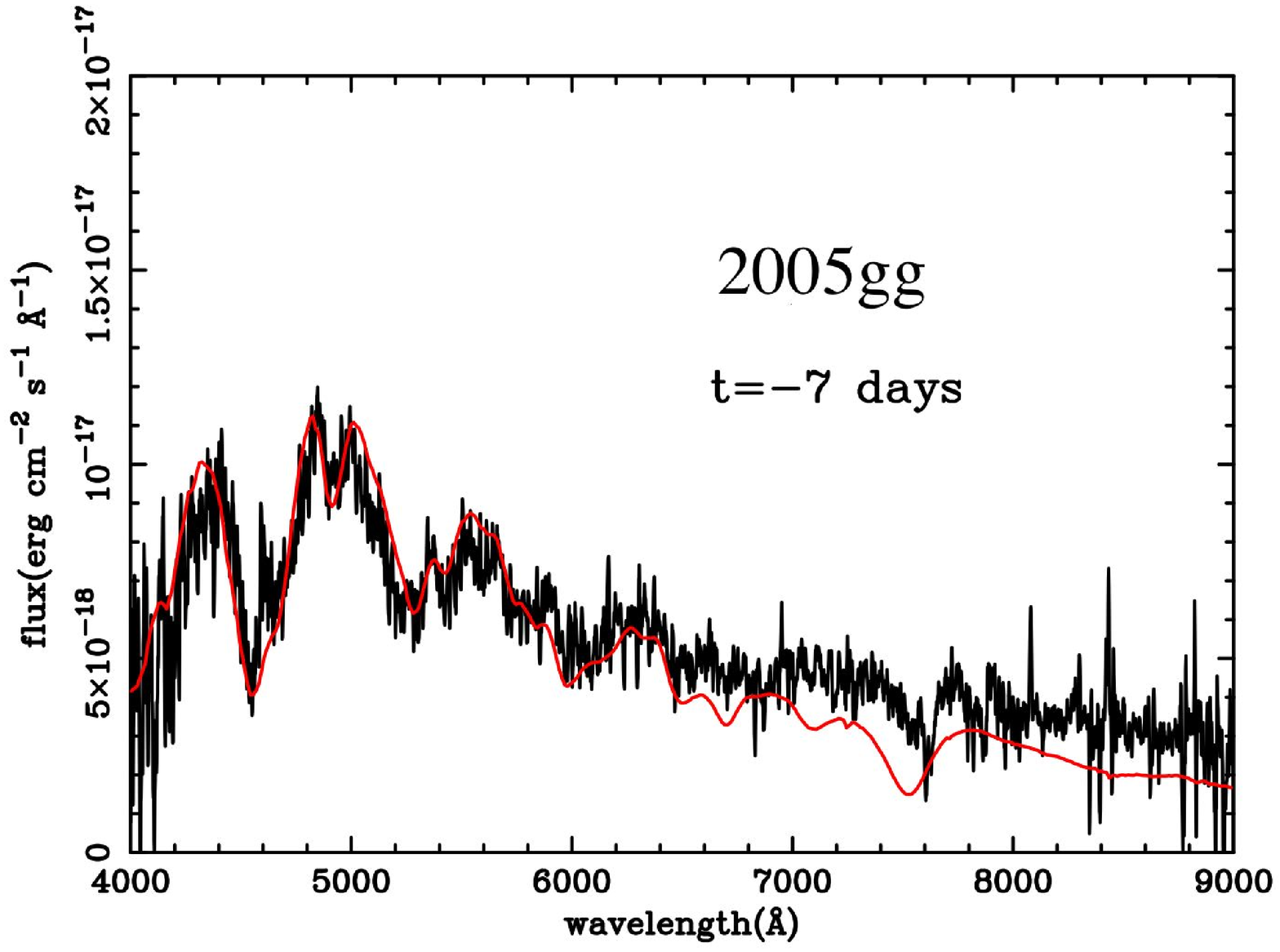}
\caption{Spectra of the SNe Ia shown in Fig. \ref{fig:lightcurve}.
{\it Left:} WHT spectrum of SN 2005ff; {\it Right:} Subaru 
spectrum of SN 2005gg. Galaxy light has not been subtracted. Red curves 
denote template SN Ia spectra at the indicated epochs relative to peak 
light. Abscissa shows observed wavelength.
}
\label{fig:spectra}
\end{figure}

\begin{figure}[!ht]
\centering
\includegraphics[width=3.5in]{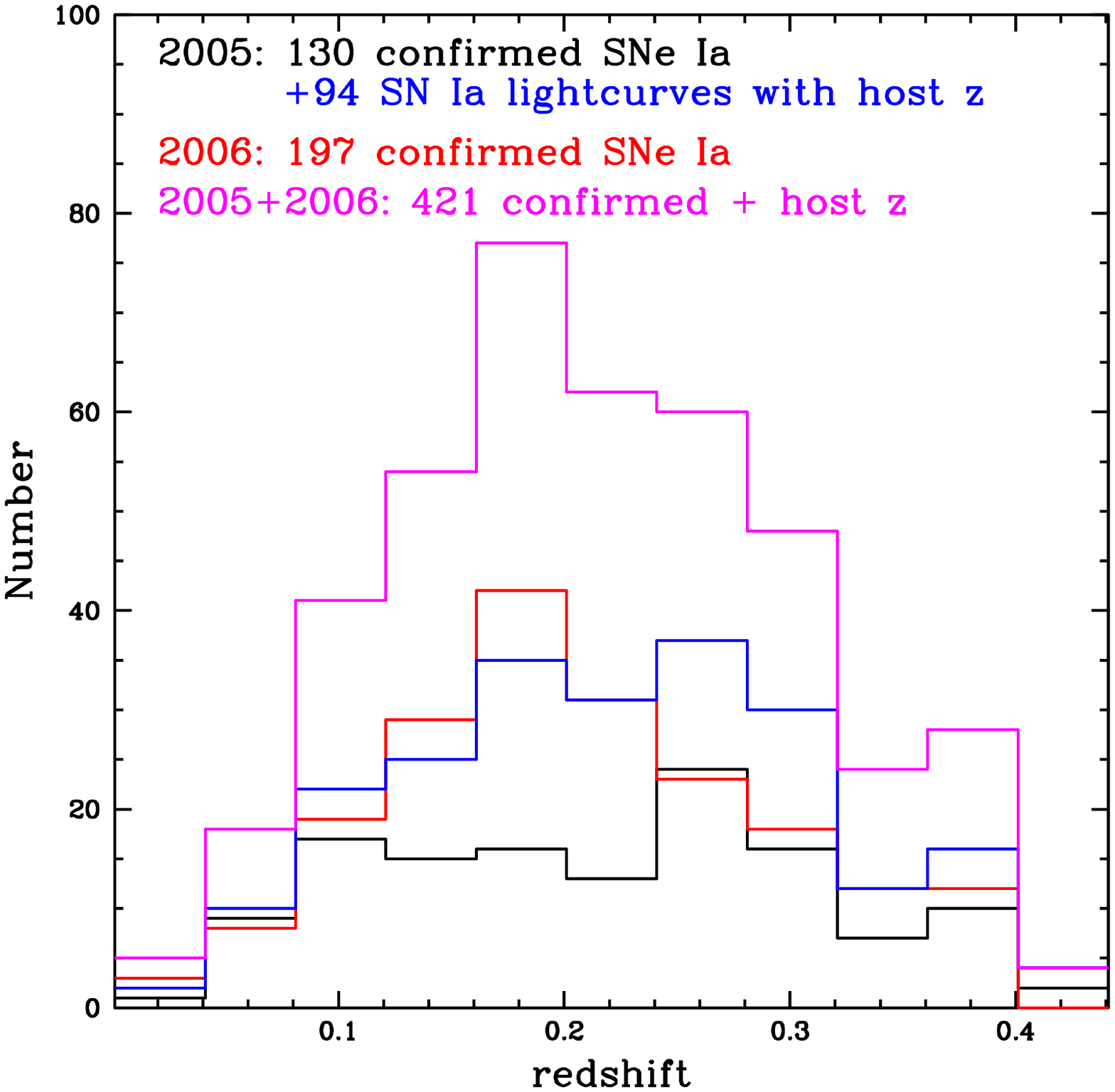}
\caption{Histogram of redshifts of spectroscopically confirmed 
SNe Ia for the 2005 (short-dashed) and 2006 (dotted) seasons. Long-dashed 
histogram
shows the 2005 confirmed sample plus 94 objects with SN Ia light curves 
and subsequent host-galaxy redshift measurements. Solid histogram 
shows the sum of these three distributions. 
}
\label{fig:spec_hist}
\end{figure}

Spectra are analyzed (Zheng et al. 2007) to determine SN type, and 
the redshift is determined from either the SN or host-galaxy 
spectrum (or both) using standard cross-correlation techniques. 
Spectra that are 
consistent with those of SNe Ia 
but for which the identification is not highly confident are denoted 
probable SNe Ia. Two spectra from the 2005 season are 
shown in Figure \ref{fig:spectra}.

The early light-curve fitting and target selection 
algorithm described above in \S \ref{sec:select} has proven to be 
very efficient at photometrically identifying SNe Ia. 
Approximately 90\% of the objects targeted as SN Ia candidates after two or 
more epochs of photometry indeed turned out to be SNe Ia; about 
86\% of the targets were spectroscopically confirmed, while the other  
4\% have high signal-to-noise, multi-band light curves that leave no doubt 
that they are SNe Ia. The remaining spectra (about 10\%) 
are mostly unclassifiable, 
often as a result of marginal observing conditions, but also include 
a very small percentage of active galactic nuclei (AGN) and 
core-collapse SNe ($\sim 2$\%). 
In addition to SNe Ia, 
we have also targeted a sample of bright ($r < 20$~mag) 
core-collapse SN candidates.  Although the colors of these events 
could be confused with those of an AGN, we generally avoid candidates 
that appear within $\sim 0.5\arcsec$ of the host galaxy nucleus.  
The confirmation 
efficiency for these events is  $\sim 90$\%, comparable to that of the 
SNe Ia.

Figure \ref{fig:spec_hist} shows the redshift distributions for the 
spectroscopically confirmed SNe Ia from the 2005 (black) and 2006 (red) 
seasons. 
For 2005, there were 130 spectroscopically confirmed and 16
spectroscopically probable SNe Ia; for 2006, the corresponding 
numbers were 197 and 14. The major difference between the 
two seasons is the large 
increase in the spectroscopic yield in the redshift 
range $0.12<z<0.24$; this is attributable to changes 
in targeting strategy (we placed less emphasis on the highest-redshift 
candidates in 2006) and to the large increase in $3-4$ m 
telescope resources for spectroscopy, as shown in Table \ref{tab1}. 
The blue histogram shows the 2005 confirmed SN Ia sample augmented by 
a set of 94 host-galaxy redshifts for objects with SN Ia-like light 
curves (see below), and the magenta histogram shows the redshift distribution  
for the combined sample of 421 SNe Ia to date. The mean redshift for 
the combined sample is $z=0.22$. 

\begin{figure}[!ht]
\centering
\includegraphics[width=3.5in]{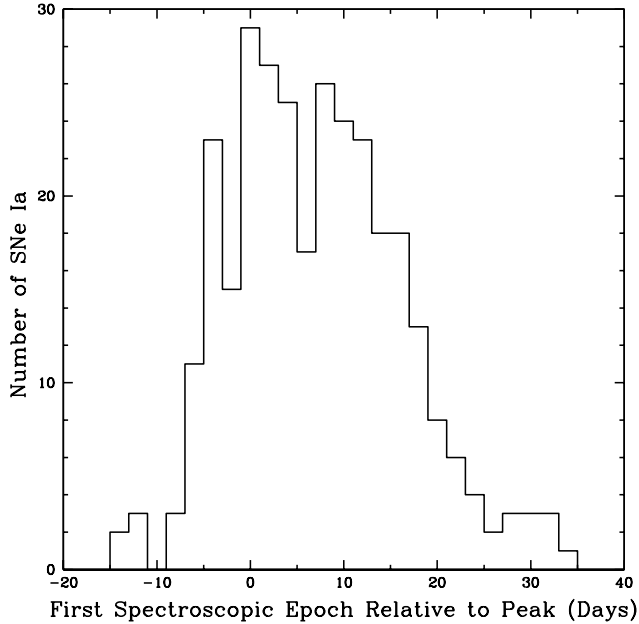}
\caption{Distribution of epoch relative to $g$-band peak light 
of first spectroscopic observations  
for the spectroscopically confirmed SNe Ia from the 2005 and 2006 seasons. 
The epoch of peak light is estimated from light-curve fits to the 
on-mountain photometry. 
}
\label{fig:spec_epoch}
\end{figure}

Figure \ref{fig:spec_epoch} shows the distribution of the epoch 
of first spectroscopic observation relative to peak light in the $g$ band 
for the spectroscopically confirmed SNe Ia from 2005 and 2006, 
where the latter is determined from light-curve fits to the 
on-mountain photometry (Sako et al. 2007). Two-thirds 
of the first-epoch spectra 
were taken within $\pm 10$ days of peak light, when a SN Ia is within 
roughly 1 mag of maximum, and 28\% were taken 
before peak light. 

Since spectroscopic resources are limited, we cannot target all 
SN Ia candidates, even all those with good-quality 
light curves: targets listed at 
low priority may not make it onto the queue of any of the  
telescopes or may not be observed before fading, due to poor weather, etc.
To gauge the resulting 
completeness, after each season we reprocess a large subset of 
the SN candidates. For the 2005 season, to find additional SNe Ia, 
we first selected  
all photometric SN candidates detected at two or more epochs 
by the on-mountain photometry pipeline. Those for which the 
best-fit light-curve template was a SN Ia were processed 
through the final scene-modeling photometry pipeline; in many cases, 
additional detection epochs were thereby recovered.  
For candidates with resulting high-quality SN 
Ia-like light curves which had not been spectroscopically confirmed, we 
carry out host-galaxy spectroscopy, as resources permit, to determine 
the redshift and measure host-galaxy properties. To date, we have 
obtained host spectra and redshifts 
for 81 such photometric SN Ia candidates from 
the 2005 season, primarily with HET. An additional 13 photometric SN Ia 
candidates have host spectra from the SDSS redshift survey. 
The redshift distribution for 
this combined photometric sample, added to the spectroscopically 
confirmed 2005 supernovae, is shown as the blue histogram in 
Figure \ref{fig:spec_hist}. We see that these objects largely 
fill in the gap at $z \simeq 0.2$ in the 2005 spectroscopic sample. 

\begin{figure}[!ht]
\centering
\includegraphics[width=5.5in]{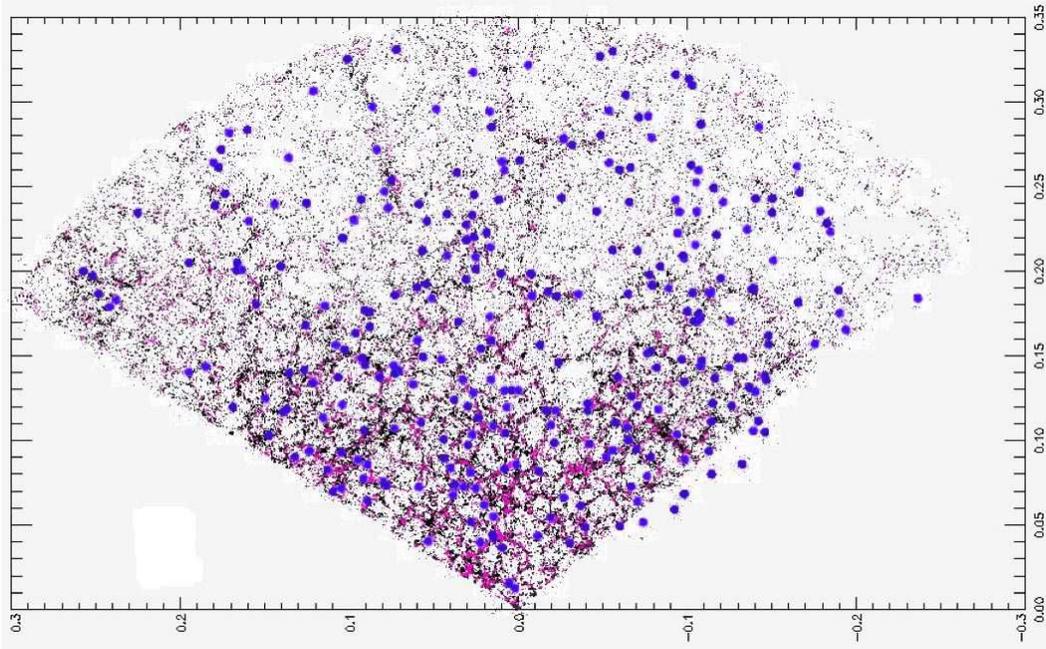}
\caption{Distribution of confirmed SDSS-II SNe Ia from 2005 and 
2006 in RA and redshift (large blue points), 
superposed on the distribution of 
galaxies with redshifts measured by the SDSS (small points in black 
and magenta).
The SN survey extends slightly beyond the RA limits for 
the redshift survey, leading to the handful of SNe that appear 
``out of bounds''.  
}
\label{fig:pie}
\end{figure}

The distribution in RA and redshift for the spectroscopically 
confirmed SNe Ia is shown in Figure \ref{fig:pie}. The SNe Ia 
are shown (large purple points) superposed on the 
distribution of galaxies with measured redshifts from the 
SDSS (black and red points, with color indicating whether the 
galaxy is at positive or negative declination within stripe 82).

\begin{figure}[!ht]
\centering
\includegraphics[width=3.5in]{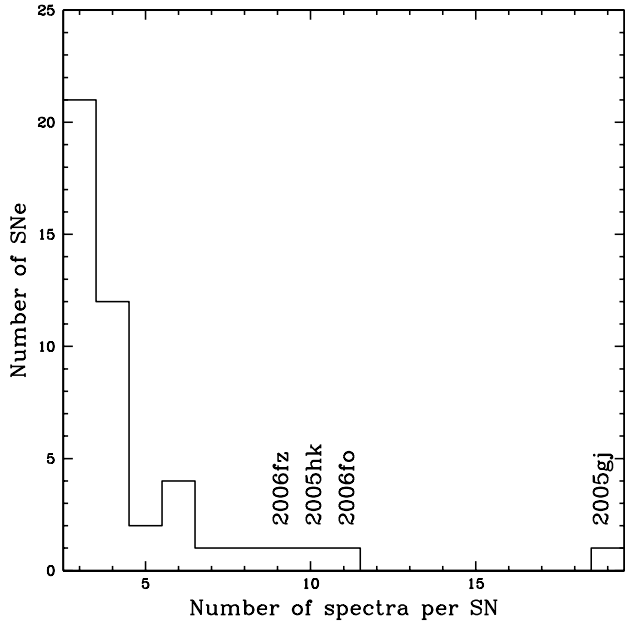}
\caption{Number of supernovae vs. number of spectroscopic 
epochs per supernova for confirmed SNe with at least three spectroscopic 
epochs, from the 2005 and 2006 seasons combined. IAU designations 
for those with the most epochs of spectroscopy are indicated. Not shown 
are the 44 confirmed SNe with two epochs of spectroscopy. SN 2006fo is a SN 
Ib/c at $z=0.02$; SN 2006fz is a SN Ia at $z=0.11$.
}
\label{fig:nspec}
\end{figure}

A subsample of the confirmed SNe are targeted for several or more 
epochs of spectroscopy. These include peculiar or otherwise interesting 
SNe Ia, 
e.g., SN 2005hk (Phillips et al. 2007) and SN 2005gj (Prieto et al. 2007), 
for which spectral evolution can inform physical modeling. 
In addition, normal SNe Ia with  
several epochs of spectroscopy will be used to explore development of 
improved K-correction models and to study correlations between 
spectroscopic evolution and photometric properties. 
Figure \ref{fig:nspec} shows a histogram of the number of 
spectroscopic epochs for SNe with at least three such epochs. 

In addition to SDSS photometry and follow-up spectroscopy, a number 
of SDSS SNe have been imaged in SDSS filters and in other pass bands 
with other telescopes. These additional photometric data are 
taken for a variety of reasons: to provide more rapid confirmation of 
SNe at early epochs, to fill in SDSS light curves during periods 
of bad weather at APO, and to extend light curves for SNe that 
have faded beyond the SDSS detection limit or that are still 
near peak light at the end of the SDSS observing season on Nov. 30. 
Telescopes used for this purpose include the University of Hawaii 
2.2 m, the Hiltner 2.4 m at MDM, the 
New Mexico State University 1 m at APO, the ARC 3.5 m, 
the 1.8 m Vatican Advanced Technology Telescope 
at Mt. Graham, the 3.5 m WIYN 
telescope at Kitt Peak, the 1.5 m optical telescope 
at Maidanak Observatory in Uzbekistan, and the 2.5 m Isaac Newton Telescope at 
La Palma. For 2005 and 2006,  
177 confirmed SNe were imaged at least once by one of these 
telescopes. In addition, the Carnegie 
Supernova Project is obtaining densely sampled optical and 
near-infrared (NIR) imaging for many of the SDSS SNe at $z<0.08$ and 
NIR imaging near peak for a subsample of SDSS SNe at $z \simeq 0.3$. 
Given the differences in filter response for this variety of 
instruments, careful attention to color terms and calibrations 
is necessary to combine 
these data with SDSS photometry (see Holtzman et al. 2007). 

\section{Conclusion}

The SDSS-II Supernova Survey has 
completed the first two of its planned three three-month observing seasons. 
Light curves have been measured for 327 spectroscopically confirmed 
SNe Ia. Including spectroscopically probable SNe Ia, other confirmed 
SN types, and objects with SN Ia-like light curves and host-galaxy 
spectroscopy, the yield to date is 496. The Fall 2007 season should 
increase the sample size by about 50\%.

\acknowledgements

We acknowledge Christopher Stubbs for his early work developing the 
case for this project and Phil Pinto for sharing 
his supernova Monte Carlo simulation code during the planning stages. 
We thank Pinto, Nick Suntzeff, Saul Perlmutter, Chris Smith, and 
Chris Pritchet 
for valuable advice in reviewing the SDSS-II SN Survey in the concept stage.

Funding for the creation and distribution of the SDSS and SDSS-II
has been provided by the Alfred P. Sloan Foundation,
the Participating Institutions,
the National Science Foundation,
the U.S. Department of Energy,
the National Aeronautics and Space Administration,
the Japanese Monbukagakusho,
the Max Planck Society, and the Higher Education Funding Council for England.
The SDSS Web site \hbox{is {\tt http://www.sdss.org/}.}

The SDSS is managed by the Astrophysical Research Consortium
for the Participating Institutions.  The Participating Institutions are
the American Museum of Natural History,
Astrophysical Institute Potsdam,
University of Basel,
Cambridge University,
Case Western Reserve University,
University of Chicago,
Drexel University,
Fermilab,
the Institute for Advanced Study,
the Japan Participation Group,
Johns Hopkins University,
the Joint Institute for Nuclear Astrophysics,
the Kavli Institute for Particle Astrophysics and Cosmology,
the Korean Scientist Group,
the Chinese Academy of Sciences (LAMOST),
Los Alamos National Laboratory,
the Max-Planck-Institute for Astronomy (MPA),
the Max-Planck-Institute for Astrophysics (MPiA), 
New Mexico State University, 
Ohio State University,
University of Pittsburgh,
University of Portsmouth,
Princeton University,
the United States Naval Observatory,
and the University of Washington.

This work is based in part on observations made at the 
following telescopes.
The Hobby-Eberly Telescope (HET) is a joint project of the University of Texas
at Austin,
the Pennsylvania State University,  Stanford University,
Ludwig-Maximillians-Universit\"at M\"unchen, and Georg-August-Universit\"at
G\"ottingen.  The HET is named in honor of its principal benefactors,
William P. Hobby and Robert E. Eberly.  The Marcario Low-Resolution
Spectrograph is named for Mike Marcario of High Lonesome Optics, who
fabricated several optical elements 
for the instrument but died before its completion;
it is a joint project of the Hobby-Eberly Telescope partnership and the
Instituto de Astronom\'{\i}a de la Universidad Nacional Aut\'onoma de M\'exico.
The Apache 
Point Observatory 3.5 m telescope is owned and operated by 
the Astrophysical Research Consortium. We thank the observatory 
director, Suzanne Hawley, and site manager, Bruce Gillespie, for 
their support of this project.
The Subaru Telescope is operated by the National 
Astronomical Observatory of Japan. The William Herschel 
Telescope is operated by the 
Isaac Newton Group, and the Nordic Optical Telescope is 
operated jointly by Denmark, Finland, Iceland, Norway, 
and Sweden, both on the island of La Palma
in the Spanish Observatorio del Roque 
de los Muchachos of the Instituto de Astrofisica de 
Canarias. Observations at the ESO New Technology Telescope at La Silla
Observatory were made under programme IDs 77.A-0437, 78.A-0325, and 
79.A-0715.
Kitt Peak National Observatory, National Optical 
Astronomy Observatory, is operated by the Association of 
Universities for Research in Astronomy, Inc. (AURA) under 
cooperative agreement with the National Science Foundation. 
The WIYN Observatory is a joint facility of the University of 
Wisconsin-Madison, Indiana University, Yale University, and 
the National Optical Astronomy Observatories.
The W.M. Keck Observatory is operated as a scientific partnership 
among the California Institute of Technology, the University of 
California, and the National Aeronautics and Space Administration. The 
Observatory was made possible by the generous financial support of the 
W. M. Keck Foundation. 
The South African Large Telescope of the South African Astronomical 
Observatory is operated by a partnership between the National 
Research Foundation of South Africa, Nicolaus Copernicus Astronomical 
Center of the Polish Academy of Sciences, the Hobby-Eberly Telescope 
Board, Rutgers University, Georg-August-Universit\"at G\"ottingen, 
University of Wisconsin-Madison, University of Canterbury, University 
of North Carolina-Chapel Hill, Dartmouth College, Carnegie Mellon 
University, and the United Kingdom SALT consortium.

\end{document}